\documentclass[pra, aps, twocolumn, floatfix, showpacs]{revtex4-1}
\usepackage{graphicx, amsmath, amssymb, times}

\begin{document}
\title{Topological phase transitions on a triangular optical lattice with non-Abelian gauge fields}
\author{M. Iskin}
\affiliation{
Department of Physics, Ko\c c University, Rumelifeneri Yolu, 34450 Sar{\i}yer, Istanbul, Turkey.
}
\date{\today}
\begin{abstract}
We study the mean-field BCS-BEC evolution of a uniform Fermi gas on a 
single-band triangular lattice, and construct its ground-state phase diagrams, 
showing a wealth of topological quantum phase transitions between gapped 
and gapless superfluids that are induced by the interplay of an out-of-plane 
Zeeman field and a generic non-Abelian gauge field. 
\end{abstract}
\maketitle

\section{Introduction}
\label{sec:intro}

The intriguing possibility of superfluid (SF) phase transitions from topology in 
momentum $(\mathbf{k})$ space has long been of interest not only to the 
condensed-matter but also to the cold atom and molecular physics communities 
in the broad contexts of nodal superconductors ($d$-wave symmetry for 
high-$T_c$ materials), nodal SFs ($p$-wave symmetries for liquid $^3$He 
and single-component Fermi gases), and population-imbalanced SFs 
($s$-wave symmetry for two-component Fermi gases)~\cite{volovik07}. 
Analogous to the Lifshitz transition in metals~\cite{lifshitz63}, these topological 
phase transitions are solely associated with the appearance or disappearance 
of $\mathbf{k}$-space regions with zero excitation energies, where the 
symmetry of the SF order parameter remains unchanged in sharp contrast 
to the Landau's classification of ordinary phase transitions. Not only such 
changes naturally cause a dramatic rearrangement of particles in 
$\mathbf{k}$ space, and therefore, are readily seen in their 
$\mathbf{k}$-resolved distribution and/or spectral function, but they also 
leave non-analytic signatures in the thermodynamic properties 
of the system~\cite{iskin06}. 

To observe and study topological phase transitions, one requires to have a 
reliable knob over either the density of particles or both the strength and 
symmetry of the inter-particle interactions~\cite{read00}. Since such controls 
are either very limited or not yet possible in condensed-matter systems, 
the cold-atom systems initially thought to offer an ideal platform for realizing 
these transitions, thanks in particular to their precise-tuning capabilities 
over a wide range of laser parameters. 
However, despite all the past and ongoing attempts 
with Fermi gases across $p$-wave Feshbach 
resonances~\cite{gaebler07, inada08, fuchs08, maier10}, the short lifetimes 
of the resultant $p$-wave molecules have so far been the biggest drawback 
in this line of research, which the experimentalists yet to overcome.

On the other hand, given the recent progress in creating artificial gauge 
fields~\cite{dalibard11, galitski13}, there is a growing consensus that one 
of the most promising ways to realize a topological phase transition is to 
incorporate $s$-wave Fermi gases with spin-orbit couplings 
(SOC)~\cite{sato09}. For instance, depending on the inter-particle 
interaction, polarization, dimension, geometry, and symmetry and strength 
of SOC, it is possible to create a zoo of nodal SFs with point, line or 
surface nodes in $\mathbf{k}$ space in various numbers.
Since several groups have already succeeded in creating such setups at 
high temperatures~\cite{wang12, cheuk12, williams13, fu13}, 
there is arguably no doubt that these new systems will 
soon offer unforeseen possibilities once they are cooled below the required 
SF transition temperature. Stimulated by these experiments, there has been 
a fruitful activity on many aspects of spin-orbit coupled Fermi gases, 
but the majority of them are focused on continuum systems 
with a lack of interest in lattice ones~\cite{zhai14}. For instance, even though topological 
SFs have recently been charecterized for a square lattice with non-Abelian
gauge fields~\cite{kubasiak10, iskin16}, and tunable honeycomb lattices 
(made of two triangular sublattices) are of both ongoing experimental and 
theoretical interest~\cite{tarruell12, struck12, hauke12, jotzu14, kim13}, 
the triangular lattices themselves are almost entirely overlooked in this context.

Here, we study the BCS-BEC evolution of a spin-$1/2$ Fermi gas on a 
single-band triangular lattice, and construct its ground-state phase diagrams. 
Our primary objective is to establish that the interplay of an out-of-plane 
Zeeman field and a generic non-Abelian gauge field gives rise to a wealth of 
topological phase transitions between gapped and gapless SFs
that are accessible in atomic optical lattices. 
The rest of the paper is organized as follows. After we introduce the model 
Hamiltonian in Sec.~\ref{sec:mft}, first we discuss the effects of a generic 
non-Abelian gauge field on the single-particle problem, and then briefly summarize 
the mean-field formalism that is used for tackling the many-body problem.
In Sec.~\ref{sec:topo}, we thoroughly analyze the conditions under which the 
quasiparticle/quasihole excitation spectrum of the SF phase may vanish, and 
evaluate the corresponding changes in the underlying Chern number.
These conditions are numerically solved in Sec.~\ref{sec:sce} together with 
the self-consistency equations, where we construct the ground-state phase 
diagrams as a function of particle filling and SOC for a wide range of 
polarizations and interactions. The paper ends with a briery summary of our 
conclusions and an outlook given in Sec.~\ref{sec:conc}.

\section{Theoretical Model}
\label{sec:mft}

In this paper, we consider a spin-$1/2$ Fermi gas on a triangular lattice, and 
study the effects of the following non-Abelian gauge field
$
\mathbf{A}=(\alpha \sigma_y, -\beta\sigma_x)
$ 
on the ground state SF phases, where $\{ \alpha, \beta \} \ge 0$ characterize the SOC, 
and $\sigma_x$ and $\sigma_y$ are the Pauli-spin matrices~\cite{kubasiak10, iskin16}. 
We take the gauge field into account via the Peierls substitution, under which 
the tunneling of atoms from site $i$ to $j$ is described by the hopping Hamiltonian
$
H_0 = - \sum_{\sigma \sigma' ij} c_{\sigma' j}^\dagger t_{j i}^{\sigma' \sigma} c_{\sigma i}.
$
Here, we only allow nearest-neighbor hoppings with
$
t_{j i}^{\sigma' \sigma} = t e^{-i \int_\mathbf{r_i}^\mathbf{r_j} \mathbf{A} \cdot d\mathbf{r}},
$ 
where $t \ge 0$ is its amplitude, and $\mathbf{r_i}$ is the position of site $i$.
The lattice spacing $a$ is set to unity in this paper. 
Using the Fourier series expansion of the annihilation operator 
$
c_{\sigma i} = ( 1/\sqrt{M}) \sum_\mathbf{k} e^{{\rm i} \mathbf{k} \cdot \mathbf{r_i}} c_{\sigma \mathbf{k}}
$
and its Hermitian conjugate in momentum $\mathbf{k} = (k_x, k_y)$ space, 
where $M \to \infty$ is the number of lattice sites in the system, the hopping 
Hamiltonian can be written as
$
H_0 = \sum_{\mathbf{k}} \psi_{\mathbf{k}}^\dagger 
(\epsilon_\mathbf{k} \sigma_0
+ \mathbf{S}_\mathbf{k} \cdot \vec{\sigma})
\psi_{\mathbf{k}}.
$
Here, the spinor
$
\psi_{\mathbf{k}}^\dagger = (c_{\uparrow \mathbf{k}}^\dagger, c_{\downarrow \mathbf{k}}^\dagger)
$
denotes the creation operators, $\epsilon_\mathbf{k}$ is the energy dispersion, 
$\sigma_0$ is the identity matrix, $\mathbf{S}_\mathbf{k} = (S_\mathbf{k}^x , S_\mathbf{k}^y, 0)$ 
is the SOC, and $\vec{\sigma} = (\sigma_x, \sigma_y, \sigma_z)$ is a vector
of spin matrices. For a triangular crystal lattice with primitive unit vectors 
$\mathbf{a_1} = (1,0)$ and $\mathbf{a_2} = (1/2,\sqrt{3}/2)$, we find
\begin{align}
\label{eqn:ek.t}
\epsilon_\mathbf{k} &= - 2t \cos \alpha \cos k_x - 4t \cos \gamma \cos \frac{k_x}{2} \cos \frac{\sqrt{3} k_y}{2}, \\
\label{eqn:skx.t}
S_\mathbf{k}^x &= - 2 \sqrt{3} t \beta \frac{\sin \gamma} {\gamma} \cos \frac{k_x}{2} \sin \frac{\sqrt{3}k_y}{2}, \\
\label{eqn:sky.t}
S_\mathbf{k}^y &= 2t \sin \alpha \sin k_x + 2 t \alpha \frac{\sin \gamma} {\gamma} \sin \frac{k_x}{2} \cos \frac{\sqrt{3}k_y}{2},
\end{align}
where $\gamma = \sqrt{\alpha^2+3\beta^2}/2$.
Note that the reciprocal of a triangular lattice is a hexagonal lattice in 
$\mathbf{k}$ space with primitive unit vectors $\mathbf{b_1} = (2\pi,-2\pi/\sqrt{3})$ 
and $\mathbf{b_2} = (0, 4\pi/\sqrt{3})$, and therefore, the first BZ is bounded 
by $|k_y| = 2\pi/\sqrt{3}$ for $|k_x| \le 2\pi/3$,
$k_y = \pm (\sqrt{3} k_x - 4\pi/\sqrt{3})$ for $2\pi/3 \le k_x \le 4\pi/3$
and $k_y = \pm (\sqrt{3} k_x + 4\pi/\sqrt{3})$ for $-4\pi/3 \le k_x \le -2\pi/3$.
In addition, since the area of the first BZ is $8\pi^2/\sqrt{3}$, we evaluate the 
sums $\sum_\mathbf{k}$ by converting them into integrals 
$
[M \sqrt{3}/(8\pi^2)] \int_\textrm{BZ} d^2\mathbf{k}
$
in our numerics.

\begin{figure}[htb]
\vskip -0.25cm
\centerline{\scalebox{0.29}{\includegraphics{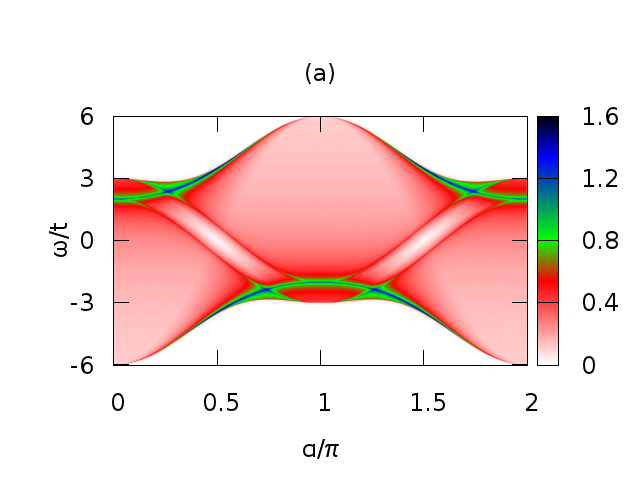}}}
\vskip -0.25cm
\centerline{\scalebox{0.29}{\includegraphics{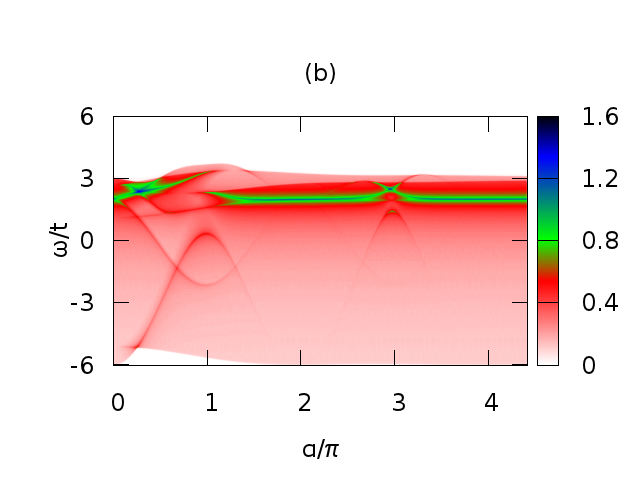}}}
\vskip -0.25cm
\centerline{\scalebox{0.29}{\includegraphics{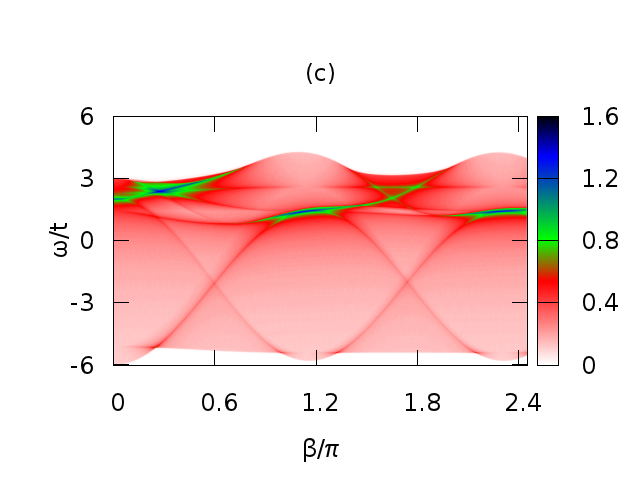}}}
\vskip -0.25cm
\centerline{\scalebox{0.29}{\includegraphics{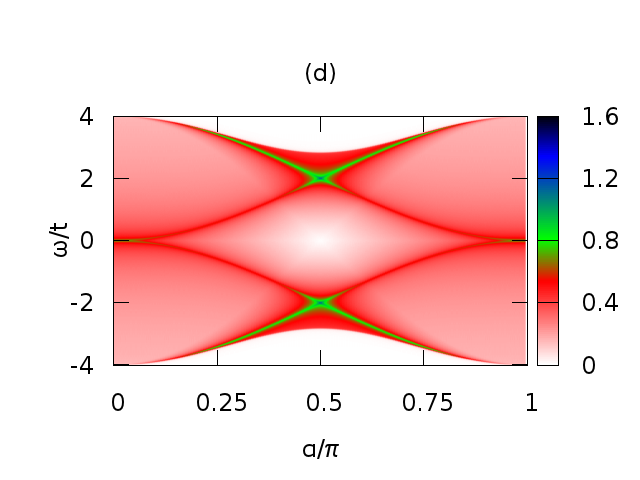}}}
\caption{\label{fig:dos} (Color online)
The total DoS $t D (\omega)$ is shown as a function of energy $\omega/t$ and 
SOC $(\alpha, \beta)$ for (a) $\alpha = \beta$, (b) $\beta = \pi/4$,  
(c) $\alpha = \pi/4$, and (d) $\alpha = \beta$ (square lattice). 
}
\end{figure}

The total single-particle density-of-states (DoS) for the hopping Hamiltonian can 
be written as $D (\omega) = D_+ (\omega) + D_-(\omega)$, where
$
D_\pm (\omega) = (1/M) \sum_\mathbf{k} \delta(\omega - \epsilon_\mathbf{k} \mp |S_\mathbf{k}|)
$
with $\delta(x)$ the Dirac-delta function. In Fig.~\ref{fig:dos}, we show colored 
maps of $t D (\omega)$ as a function of $\omega/t$ and $(\alpha, \beta)$, where 
$\pi \delta(x) \to \eta/(x^2 + \eta^2)$ is implemented in our numerics with a small 
broadening $\eta = 0.01 t$. First of all, after summing over $\mathbf{k}$, since the 
SOC can be gauged away from the single-particle problem in the limits of $\alpha \to 0$ 
or $\beta \to 0$, $D(\omega)$ approach to the no-SOC value in all figures, 
showing a sharp peak at $\omega = 2t$. This is also 
the reason behind the somewhat featureless structure of Fig.~\ref{fig:dos}(b) where 
$\beta$ is small. Furthermore, for symmetric SOCs with $\alpha = \beta$, it can 
analytically be shown that the DoS has
$
D(\omega, \alpha) = D(\omega, - \alpha),
$
$
D(\omega, \pi + \alpha) = D(\omega, \pi - \alpha)
$
and
$
D(\omega, \alpha) = D(-\omega, \pi - \alpha)
$
symmetries for any $\alpha$, leading to $D(\omega = 0, \alpha = i \pi/2) = 0$ for 
any integer $i$. These symmetries and the resultant gaps are clearly illustrated 
in our numerics shown in Fig.~\ref{fig:dos}(a), and they play important roles in 
understanding the resultant ground-state SF phases as discussed below in 
Sec.~\ref{sec:sce}. Thus, in sharp contrast to the continuum systems where 
the low-energy DoS increases with increasing SOC, we show that the DoS has 
a much richer dependence on energy and SOC on a triangular lattice.
For completeness, typical DoS data are illustrated in Figs.~\ref{fig:dos}(b) 
and~\ref{fig:dos}(c) for asymmetric SOCs with $\alpha \ne \beta \ne 0$,
showing no particular symmetry in general. In comparison to Fig.~\ref{fig:dos}(a), 
we also show the analogous DoS dependence for a nearest-neighbor square lattice 
in Fig.~\ref{fig:dos}(d), manifesting the particle-hole symmetry around 
$\omega = 0$ for any $\alpha$.

Since our primary objective in this paper is to characterize distinct SF phases 
of $\uparrow$ and $\downarrow$ fermions in the presence of on-site attractive 
interactions in between, we introduce a complex parameter
$
\Delta_i = g \langle c_{\uparrow i} c_{\downarrow i} \rangle,
$
which describes the local SF order within the BCS mean-field description, 
where $g \ge 0$ is the strength of the interaction, and $\langle \cdots \rangle$ 
is a thermal average. Finally, including a possible out-of-plane Zeeman field 
$h$, the total mean-field Hamiltonian can be compactly written as~\cite{iskin16}
$
H = M |\Delta|^2/g + \sum_\mathbf{k} \xi_\mathbf{k} + (1/2) \sum_\mathbf{k} \Psi_\mathbf{k}^\dagger H_\mathbf{k} \Psi_\mathbf{k},
$
where the operator
$
\Psi_{\mathbf{k}}^\dagger = (c_{\uparrow \mathbf{k}}^\dagger, 
c_{\downarrow \mathbf{k}}^\dagger,  c_{\uparrow, -\mathbf{k}}, c_{\downarrow, -\mathbf{k}})
$
denotes the creation and annihilation operators collectively,
the matrix
\begin{align}
H_\mathbf{k} = \left( \begin{array}{cccc}
\xi_\mathbf{k}-h & S_\mathbf{k}^\perp & 0 & \Delta \\
S_\mathbf{k}^{\perp *} & \xi_\mathbf{k} + h& -\Delta & 0  \\
0 & -\Delta^* & -\xi_\mathbf{k} + h& S_\mathbf{k}^{\perp *} \\
\Delta^* & 0 &  S_\mathbf{k}^\perp & -\xi_\mathbf{k} - h
\end{array} \right)
\label{eqn:ham.k}
\end{align}
is the Hamiltonian density,
$
\xi_\mathbf{k} = \epsilon_\mathbf{k} - \mu
$
with $\mu$ the chemical potential,
$
S_\mathbf{k}^\perp = S_\mathbf{k}^x - {\rm i} S_\mathbf{k}^y
$
is the SOC, and the SF order parameter
$
\Delta = g \sum_\mathbf{k} \langle c_{\uparrow \mathbf{k}} c_{\downarrow, -\mathbf{k}} \rangle
$
is uniform in $\mathbf{k}$ space. The eigenvalues $E_{\lambda \mathbf{k}}$ of the 
Hamiltonian matrix with $\lambda = \lbrace 1,2,3,4 \rbrace$ are simply given by
$
E_{\lambda \mathbf{k}} = s_\lambda \sqrt{\xi_\mathbf{k}^2+h^2+|\Delta|^2+|S_\mathbf{k}^\perp|^2 + 2 p_\lambda A_\mathbf{k}},
$
corresponding to the quasiparticle ($s_{1,3} = p_{3,4} = +1$) and quasihole ($p_{1,2} = s_{2,4} = -1$) 
excitation energies of the system, where 
$
A_\mathbf{k} = \sqrt{(\xi_\mathbf{k}^2 + |\Delta|^2)h^2 + |S_\mathbf{k}^\perp|^2 \xi_\mathbf{k}^2}.
$
In terms of $E_{\lambda \mathbf{k}}$, the self-consistency equations can be written as~\cite{iskin16}
\begin{align}
\label{eqn:gap}
-M \frac{|\Delta|}{g} &= \frac{1}{4} \sum_{\lambda \mathbf{k}} \frac{\partial E_{\lambda \mathbf{k}}}{\partial |\Delta|} f(E_{\lambda \mathbf{k}}), \\
\label{eqn:ntot}
N_\uparrow + N_\downarrow &= \frac{1}{4} \sum_{\lambda \mathbf{k}} \left[1 - 2\frac{\partial E_{\lambda \mathbf{k}}}{\partial \mu} f(E_{\lambda \mathbf{k}}) \right], \\
\label{eqn:ndif}
N_\downarrow - N_\uparrow &= \frac{1}{2} \sum_{\lambda \mathbf{k}} \frac{\partial E_{\lambda \mathbf{k}}}{\partial h} f(E_{\lambda \mathbf{k}}),
\end{align}
where $f(x) = 1/[e^{x/(k_B T)} + 1]$ is the Fermi function with $k_B$ the Boltzmann constant 
and $T$ the temperature. Here, the derivatives are
$
\partial E_{\lambda \mathbf{k}} / \partial |\Delta| = (1 + p_\lambda h^2/A_\mathbf{k} ) |\Delta| / E_{\lambda \mathbf{k}}
$
for the order parameter,
$
\partial E_{\lambda \mathbf{k}} / \partial \mu = - [1 + p_\lambda ( h^2+|S_\mathbf{k}^\perp|^2 )/A_\mathbf{k} ] \xi_\mathbf{k} / E_{\lambda \mathbf{k}}
$
for the chemical potential, and
$
\partial E_{\lambda \mathbf{k}} / \partial h = - [1 + p_\lambda ( \xi_\mathbf{k}^2+|\Delta|^2 )/A_\mathbf{k} ] h / E_{\lambda \mathbf{k}}
$
for the Zeeman field. 
While $\mu$ determines the total number $N = N_\uparrow + N_\downarrow$ 
of atoms where $N_\sigma = \sum_i n_{\sigma i}$ with the local fermion filling
$
0 \le n_{\sigma i} = \langle c_{\sigma i}^\dagger c_{\sigma i} \rangle \le 1,
$
$h \ge 0$ determines the polarization $P = (N_\uparrow - N_\downarrow)/N \ge 0$ 
of the system which is assumed to be positive without loosing generality.
Next, we analyze $E_{\lambda \mathbf{k}}$ for gapped/gapless solutions to distinguish 
SF phases by the $\mathbf{k}$-space topology of their excitations.

\section{Topological Superfluids}
\label{sec:topo} 

\begin{figure}[htb]
\centerline{\scalebox{0.3}{\includegraphics{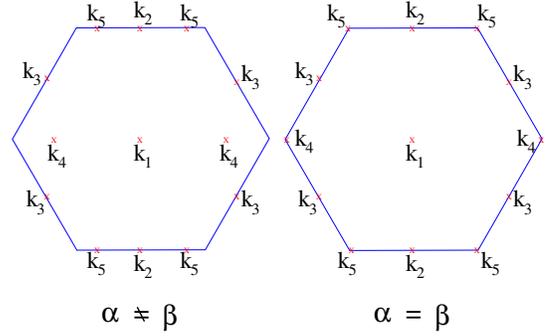}}}
\caption{\label{fig:BZ} (Color online)
The gapless $\mathbf{k}$-space points are illustrated within the first BZ. 
Note that either the set $\mathbf{k_4}$ or $\mathbf{k_5}$ but not both is 
relevant when $\alpha \ne \beta$ (see the text).
}
\end{figure}

It is clear that $E_{1 \mathbf{k}}$ and $E_{2 \mathbf{k}}$ may become gapless 
in $\mathbf{k}$ space, \textit{i.e.}, $E_{1(2) \mathbf{k_0}}$ are precisely 0 at some 
special $\mathbf{k_0}$ points satisfying the condition $|S_\mathbf{k_0}^\perp| = 0$ 
when
$
h = h_\mathbf{k_0} = \sqrt{(\epsilon_\mathbf{k_0}-\mu)^2 + |\Delta|^2}.
$
There are five sets of $\mathbf{k_0}$ points satisfying $|S_\mathbf{k_0}^\perp| = 0$:
in addition to the center $\mathbf{k_1} = (0, 0)$ of the hexagon-shaped BZ,
the two-point set $\mathbf{k_2} = (0, \pm 2\pi/\sqrt{3})$ corresponds to the midpoints of
the top and bottom edges of the BZ adding in total to one full point, 
the four-point set $\mathbf{k_3} = (\pm \pi, \pm \pi/\sqrt{3})$ corresponds to the 
midpoints of the right and left edges of the BZ adding in total to two full points, 
the two-point set $\mathbf{k_4} = (k_4^x, 0)$ is such that
$
\cos \frac{k_4^x}{2} = - \frac{\alpha}{\sin \alpha} \frac{\sin \gamma}{2 \gamma},
$
and finally the four-point set $\mathbf{k_5} = (k_5^x, \pm 2\pi/\sqrt{3})$ where 
$
\cos \frac{k_5^x}{2} = \frac{\alpha}{\sin \alpha} \frac{\sin \gamma}{2 \gamma}
$
corresponds to two half-points on the top and two half-points on the 
bottom edges adding in total to two full points.
Note that $k_5^x = k_4^x \pm 2\pi$ can only be satisfied in the first BZ 
for $\alpha = \beta$, in which case the combined set $\mathbf{k_4} = (\pm 4\pi/3, 0)$ 
and $\mathbf{k_5} = (\pm 2\pi/3, \pm 2\pi/\sqrt{3})$ corresponds to the 
six corners of the BZ adding in total to two full points.
Therefore, either the set $\mathbf{k_4}$ or $\mathbf{k_5}$ but not both
is relevant when $\alpha \ne \beta$ as illustrated in Fig.~\ref{fig:BZ}.
The corresponding energy dispersions at the location of zeros are
$\epsilon_{\mathbf{k_1}} = - 2t (\cos \alpha +2 \cos \gamma)$
for the first set,
$\epsilon_{\mathbf{k_2}} = - 2t (\cos \alpha - 2\cos \gamma)$
for the second set,
$\epsilon_{\mathbf{k_3}} = 2t \cos \alpha$
for the third set, and
$\epsilon_{\mathbf{k_4}} = \epsilon_{\mathbf{k_5}} = 2t \cos \alpha 
\left(1 - \frac{\alpha^2}{\sin^2 \alpha} \frac{\sin^2 \gamma}{2 \gamma^2} \right)
+ t \frac{\alpha}{\sin \alpha} \frac{\sin (2\gamma)}{\gamma}
$
for the remaining sets. 

It is well-known that opening or closing of a gap in $\mathbf{k}$ space gives rise to 
a topological phase transition between SFs with distinct $\mathbf{k}$-space topologies. 
In our case, these transitions are further signalled by changes in the topological 
invariant of the system~\cite{belissard95}, 
which by definition may only change due to a change in system's underlying topology. 
For instance, it can be shown that the change in CN at $h = h_\mathbf{k_0}$ 
is simply given by a sum over the Berry indices at all touching 
points~\cite{kubasiak10, iskin16}
$
\Delta {\rm CN} (h_\mathbf{k_0}) = \sum_\mathbf{k_0} {\rm sign} (Y_\mathbf{k_0})
$
where 
$
Y_\mathbf{k_0} = \left[ 
\frac{\partial S_\mathbf{k}^x}{\partial k_x} \frac{\partial S_\mathbf{k}^y}{\partial k_y}
- \frac{\partial S_\mathbf{k}^y}{\partial k_x} \frac{\partial S_\mathbf{k}^x}{\partial k_y} 
\right]_\mathbf{k_0}.
$
In particular, we find that
$
Y_\mathbf{k_1} = 6t^2 \alpha \beta
\left( \frac{\sin \alpha}{\alpha} + \frac{\sin \gamma} {2\gamma}\right) 
\frac{\sin \gamma} {\gamma}
$
is always positive and $\Delta {\rm CN} (h_\mathbf{k_1}) = +1$,
$
Y_\mathbf{k_2} = -6t^2 \alpha \beta
\left( \frac{\sin \alpha}{\alpha} - \frac{\sin \gamma} {2\gamma}\right) 
\frac{\sin \gamma} {\gamma}
$
is always negative and $\Delta {\rm CN} (h_\mathbf{k_2}) = -1$,
$
Y_\mathbf{k_3}   = - 3 t^2 \alpha \beta \frac{\sin^2 \gamma} {\gamma^2}
$
is always negative and $\Delta {\rm CN} (h_\mathbf{k_3}) = -2$, 
and 
$
Y_\mathbf{k_4} = Y_\mathbf{k_5} =  3 t^2 \alpha \beta
\left( \frac{\sin^2 \gamma} {\gamma^2} 
- \frac{\alpha^2}{\sin^2 \alpha} \frac{\sin^4 \gamma} {4\gamma^4} \right)
$
is always positive and $\Delta {\rm CN} (h_\mathbf{k_4}) = \Delta {\rm CN} (h_\mathbf{k_5}) = +2$.
Note that the total change in CN adds up to 0 for all parameters as a function
of increasing $h$. This is because since the SF phase is topologically trivial 
in the $h \to 0$ limit, the normal phase must also be topologically trivial in the $
h \gg t$ as well. In particular, when $\alpha = \beta$, these expressions reduce to
$
Y_\mathbf{k_1} = -3 Y_\mathbf{k_2} = -3 Y_\mathbf{k_3} = 4 Y_\mathbf{k_4} =
4 Y_\mathbf{k_5} =  9 t^2 \sin^2 \alpha,
$
and the corresponding energy dispersions are given by
$\epsilon_\mathbf{k_1} = -3\epsilon_\mathbf{k_2} = -3\epsilon_\mathbf{k_3}
= -2 \epsilon_\mathbf{k_4} = -2 \epsilon_\mathbf{k_5} = -6t \cos \alpha$.
Thus, since the combined set $\{ \mathbf{k_4}, \mathbf{k_5} \}$ corresponds to 
the six corners of the first BZ as illustrated in Fig.~\ref{fig:BZ}, and 
$h_\mathbf{k_2} = h_\mathbf{k_3}$ and $h_\mathbf{k_4} = h_\mathbf{k_5}$ are 
two-fold degenerate, we find $\Delta {\rm CN}(h_\mathbf{k_1}) = +1$, 
$\Delta {\rm CN}(h_\mathbf{k_{2,3}}) = -3$ and $\Delta {\rm CN}(h_\mathbf{k_{4,5}}) = +2$.
Based on this classification scheme, next we explore the phase diagrams
of the system for SF phases with distinct $\mathbf{k}$-space topologies.

\section{Self-Consistent Results}
\label{sec:sce} 

For simplicity, we restrict our numerical analysis to the ground state with symmetric 
SOCs, and solve the self-consistency Eqs.~(\ref{eqn:gap}), (\ref{eqn:ntot}) and 
(\ref{eqn:ndif}) at $T = 0$ as a function of total particle filling 
$F = (N_\uparrow + N_\downarrow)/M$ and $\alpha = \beta$. 
We construct the ground-state phase diagrams for a number of polarization 
$P = (N_\uparrow - N_\downarrow)/(N_\uparrow + N_\downarrow)$ and $g$ values. 
Note that since $0 \le N_\sigma/M \le 1$ within the single-band approximation, the 
maximum possible value of $F$ depends on $P$, \textit{i.e.}, $F_\textrm{max} = 2/(1 + |P|)$
and the majority (minority) component is a band insulator (normal) for $F \ge F_\textrm{max}$. 
In addition, since the phase diagrams are symmetric around $\pi$ with 
$\alpha \to 2\pi - \alpha$ symmetry, which is caused by the symmetry of the DoS as 
shown in Fig.~\ref{fig:dos}(a), we present the resultant phase diagrams only for 
the interval $0 \le \alpha \le \pi$. 

\begin{widetext}
\begin{center}

\begin{figure}[htb]
\vspace{-1cm}
\centerline{
\scalebox{0.345}{\includegraphics{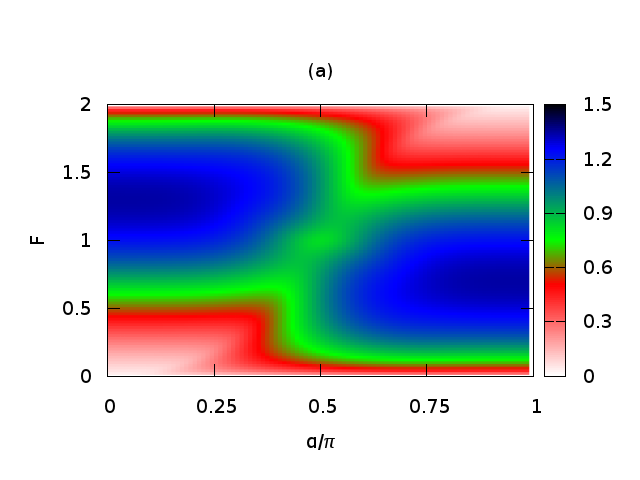}}
\scalebox{0.3}{\includegraphics{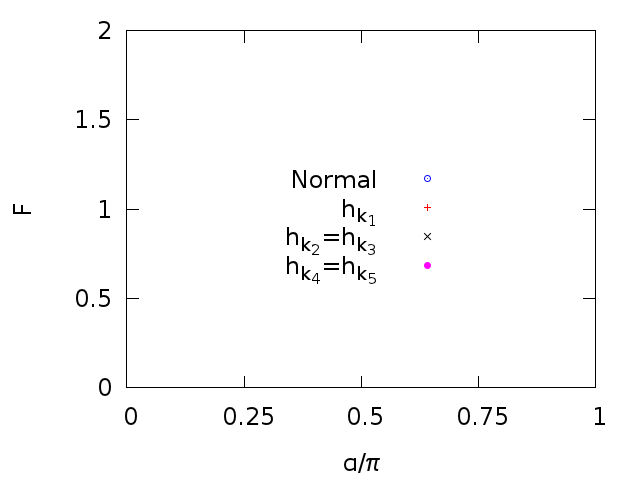}} 
}
\vspace{-0.23cm}
\centerline{
\scalebox{0.345}{\includegraphics{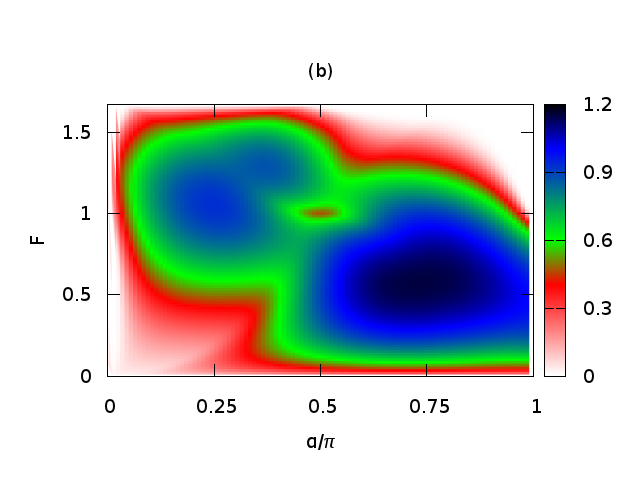}}
\scalebox{0.3}{\includegraphics{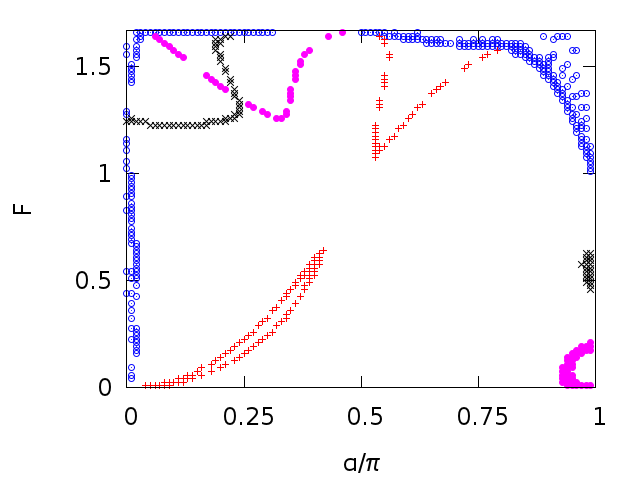}} 
}
\vspace{-0.23cm}
\centerline{
\scalebox{0.345}{\includegraphics{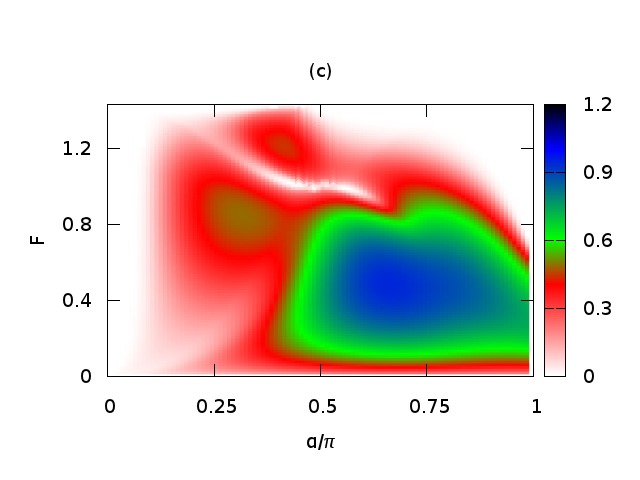}}
\scalebox{0.3}{\includegraphics{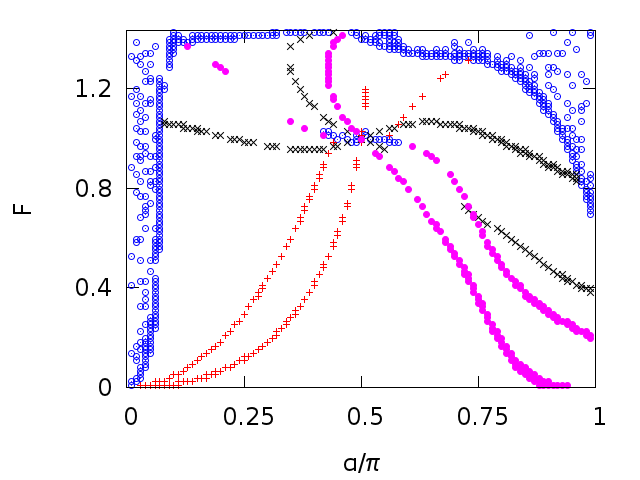}} 
}
\vspace{-0.23cm}
\centerline{
\scalebox{0.345}{\includegraphics{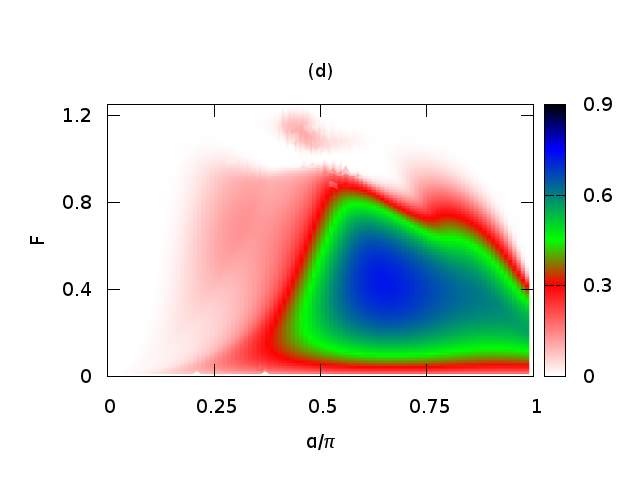}}
\scalebox{0.3}{\includegraphics{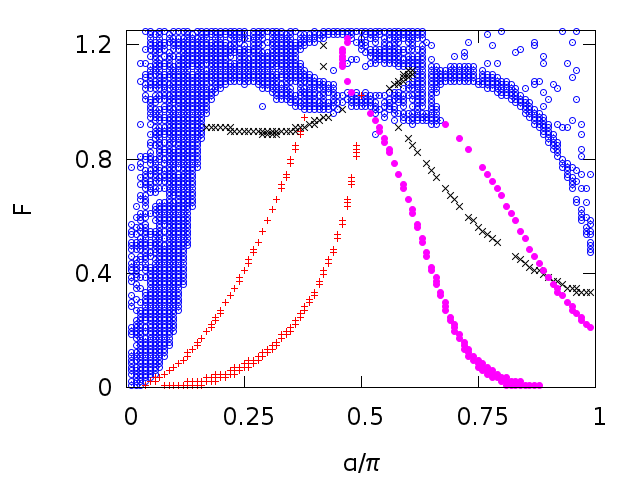}} 
}
\caption{\label{fig:P} (Color online)
The SF order parameter $|\Delta|/t$ (left) and the extracted boundaries for the 
topological quantum phase transitions (right) are shown as a function of total 
particle filling $F$ and SOC $\alpha = \beta$ for 
(a) $P = 0$, (b) $0.2$, (c) $0.4$ and (d) $0.6$, where $g = 4t$.
}
\end{figure}

\begin{figure}[htb]
\vspace{-1cm}
\centerline{
\scalebox{0.345}{\includegraphics{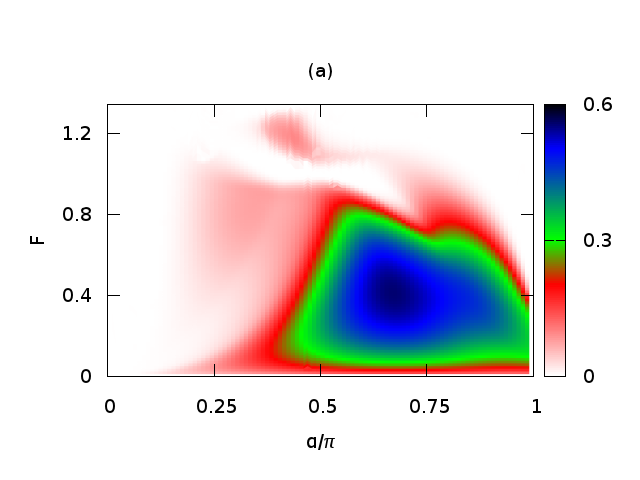}}
\scalebox{0.3}{\includegraphics{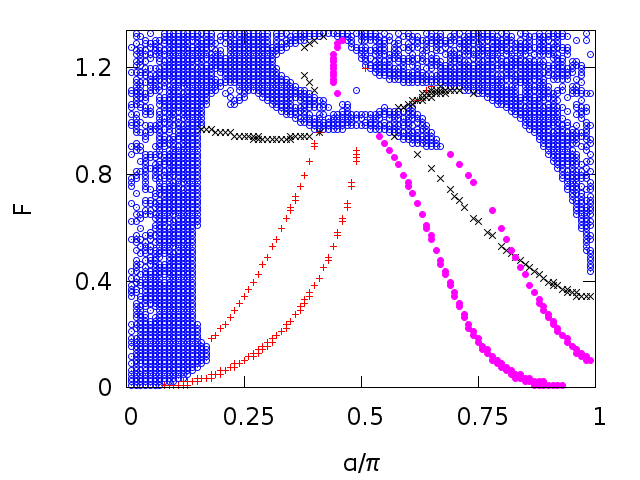}} 
}
\vspace{-0.23cm}
\centerline{
\scalebox{0.345}{\includegraphics{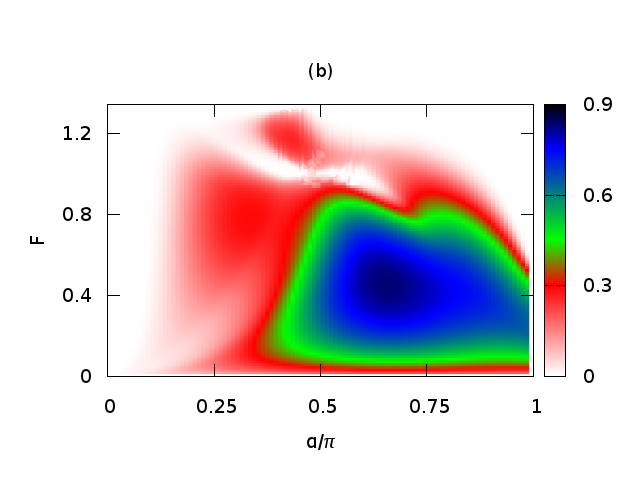}}
\scalebox{0.3}{\includegraphics{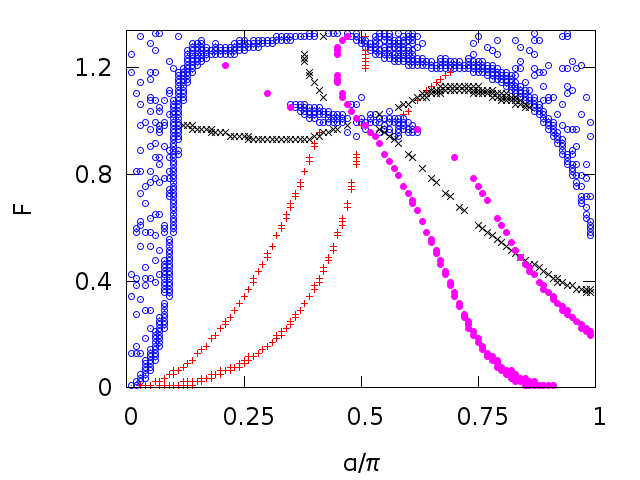}} 
}
\vspace{-0.23cm}
\centerline{
\scalebox{0.345}{\includegraphics{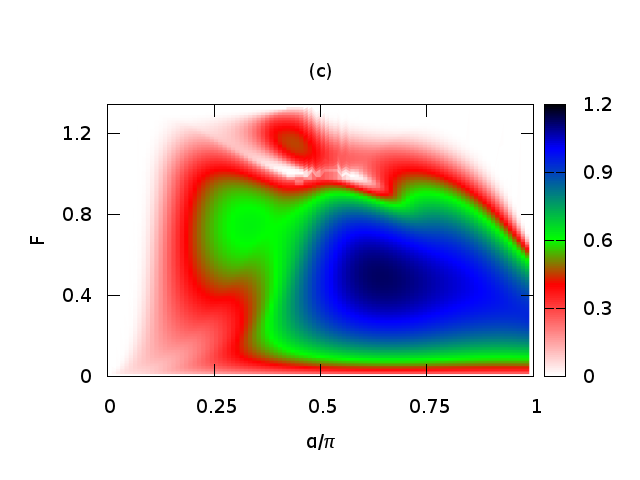}}
\scalebox{0.3}{\includegraphics{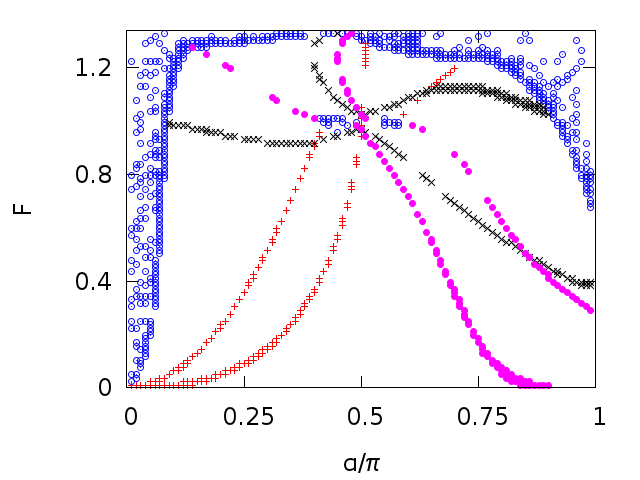}} 
}
\vspace{-0.23cm}
\centerline{
\scalebox{0.345}{\includegraphics{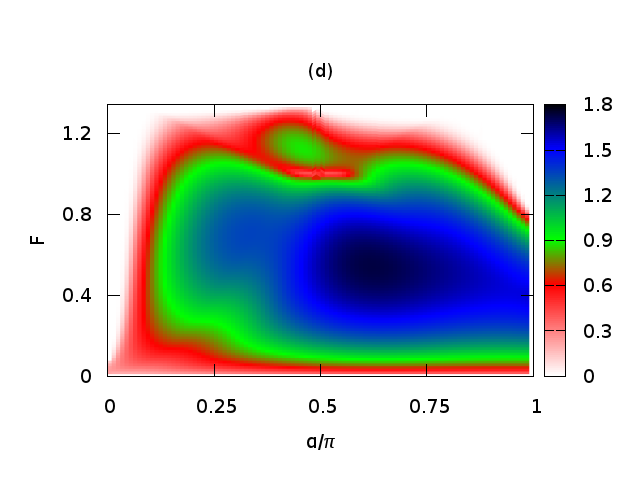}}
\scalebox{0.3}{\includegraphics{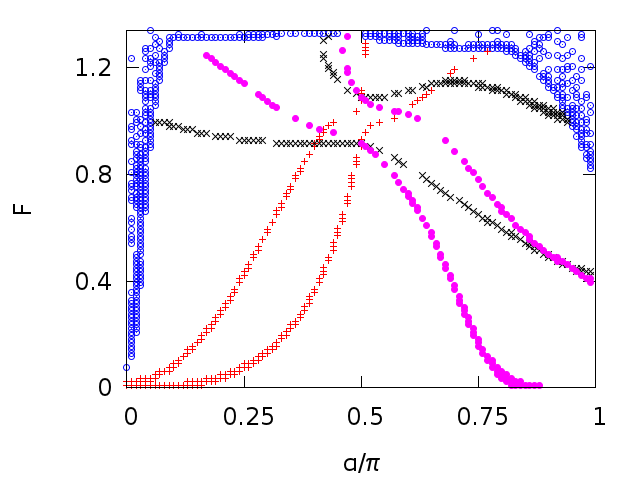}} 
}
\caption{\label{fig:g} (Color online)
The SF order parameter $|\Delta|/t$ (left) and the extracted boundaries for the 
topological quantum phase transitions (right) are shown as a function of total 
particle filling $F$ and SOC $\alpha = \beta$ for 
(a) $g = 3t$, (b) $4t$, (c) $5t$  and (d) $7t$, where $P = 0.5$.
}
\end{figure}

\end{center}
\end{widetext}

For instance, in Fig.~\ref{fig:P}, we show colored maps of $|\Delta|/t$ and the 
corresponding boundaries for the topological quantum phase transitions
when $g = 4t$ and $P = \{0, 0.2, 0.4, 0.6\}$. Here, the normal region is 
characterized by $|\Delta| \lesssim 10^{-3}t$.
In the trivial case of zero polarization shown in Fig.~\ref{fig:P}(a), the entire phase 
diagram is a SF and even though some of the low-energy features are 
smeared out by finite $g$, $|\Delta|$ has precisely the symmetry of the 
DoS shown in Fig.~\ref{fig:dos}(a), where $\omega = 0$ corresponds to 
$\mu = 0$, \textit{i.e.}, the half-filling $F = 1$. As increasing $P$ progressively 
weakens $|\Delta|$ due to the Zeeman-induced pairing mismatch between 
$\uparrow$ and $\downarrow$ fermions, not only the normal region expands 
but also more footprints of the low-energy DoS become gradually salient 
in $|\Delta|$, including the gap at $\omega = 0$ when $\alpha = \pi/2$. 
In particular, we find reentrant SF phase transitions that are interfered 
by the normal phase in the neighbourhood of this gap in 
Figs.~\ref{fig:P}(c) and~\ref{fig:P}(d). Furthermore, while the SF phase is trivially 
gapped in the entire phase diagram shown in Fig.~\ref{fig:P}(a), having a finite 
$P$ gives rise to the emergence of two phase-transition branches per each 
$\mathbf{k_i}$, satisfying the critical condition $h = h_\mathbf{k_i}$. 
These branches arise from the particle- and hole-pairing sectors, and all of 
them eventually meet at $F = 1$ and $\alpha = \pi/2$ with increasing $P$.
Since $\mu = 0$ for all parameters as long as $F = 1$, and $\epsilon_\mathbf{k_i} = 0$ 
for all $\mathbf{k_i}$ points at $\alpha = \beta = \pi/2$, all of the transition boundaries 
become degenerate precisely when $h = |\Delta|$ is simultaneously
satisfied. Therefore, the critical value of $P$ for such a crossing clearly increases 
with $g$, and it happens around $P_c \approx 0.2$ when $g = 4t$ and 
$P_c \approx 0.4$ when $g = 10t$. 

Similarly, in Fig.~\ref{fig:g}, we show colored maps of $|\Delta|/t$ and the 
corresponding boundaries for the topological quantum phase transitions
when $P = 0.5$ and $g = \{3, 4, 5, 7\}$. The intricate dependence of $|\Delta|$ 
on $F$ and $\alpha$, and the resultant reentrant SF phase transitions can again 
be traced back to the low-energy features of DoS, as they play the most 
important roles in the weakly-interacting limit where the reentrant behavior 
is most eminent for any $P$. Apart from the shrinkage of the 
normal region and gradual disappearance of the reentrant behavior due to 
enhanced pairing, one of the most notable findings in Fig.~\ref{fig:g} is that 
not only the qualitative but also the quantitative structure of the phase 
diagrams are quite robust against increasing $g$. Thus, we conclude that 
topological phase transitions with $\Delta {\rm CN} = \{ \pm1, \pm 2, \pm 3 \}$ 
are generally accessible on a triangular lattice. Having achieved our primary 
objective of constructing the ground-state phase diagrams, next we end 
this paper with a briery summary of our conclusions and an outlook.

\section{Conclusions}
\label{sec:conc} 

In summary, to describe the BCS-BEC evolution of a spin-$1/2$ Fermi gas 
that is loaded on a uniform triangular optical lattice, here we considered a 
single-band lattice Hamiltonian within the mean-field approximation for on-site pairing. 
In particular, we explored topological phase transitions between gapped 
and gapless SF phases that are induced by the interplay of an out-of-plane 
Zeeman field and a non-Abelian gauge field. These transitions are signalled by
changes in the underlying Chern number, and we found that 
$\Delta {\rm CN} = \{ \pm1, \pm 2, \pm 3 \}$ are generally accessible on a triangular 
lattice. By constructing a number of ground-state phase diagrams self-consistently 
for a wide range of parameter space, we also found reentrant SF phase transitions that 
are interfered by the normal phase, and traced their imprints to the DoS of the 
non-interacting problem. In sharp contrast to the continuum systems where
the low-energy DoS increases with increasing SOC, we showed that the DoS 
has a much richer dependence on energy and SOC on a triangular lattice, 
leading in return to an intricate dependence of the SF order parameter on 
particle filling, SOC and inter-particle interaction. Since the low-energy DoS 
plays the most important role in the weakly-interacting limit, the reentrant 
behavior is most eminent there for any polarization, and it gradually 
diminishes as the interaction gets stronger. 

Even though the topological phase transitions discussed in this paper occur in 
momentum space, and therefore, are most evident in the momentum distributions 
of atoms in time-of-flight measurements, it is well-established in the context 
of unconventional (\textit{e.g.}, $p$- or $d$-wave) nodal-SFs that such changes also 
leave non-analytic traces in the thermodynamic properties of the system 
including the compressibility, spin susceptibility, specific heat, etc. Thus, as an 
outlook, we believe it is fruitful to extend this line of research towards all sorts
of directions, including finite center-of-mass pairings, finite temperatures, 
multi-band lattices, longer-ranged hoppings/interactions, Abelian gauge fields, 
confined systems, hexagonal lattices, higher-dimensional lattices, beyond 
mean-field effects, etc. Theoretical understanding of these extensions 
in greater depth will surely have dramatic impacts for not only to the cold-atom and 
condensed-matter communities but also to the others, where the interplay of SF 
pairing and SOC are contemporary concepts offering futuristic technological applications.

\section{Acknowledgments} 
\label{sec:ack}
We gratefully acknowledge funding from T\"{U}B$\dot{\mathrm{I}}$TAK Grant No. 1001-114F232.

\end{document}